\documentclass[%
reprint, twocolumn,
superscriptaddress,
showpacs,preprintnumbers,
nofootinbib,
amsmath,amssymb, aps, pra,
]{revtex4}

\usepackage{graphicx}
\usepackage{dcolumn}
\usepackage{bm}
\usepackage{color} 
\usepackage{CJK}

\def\ra{{\rangle}}

\newcommand{\beq}{\begin{equation}}
\newcommand{\eeq}{\end{equation}}
\newcommand{\beqa}{\begin{eqnarray}}
\newcommand{\eeqa}{\end{eqnarray}}

\begin{document}
\title{Vibrational mode multiplexing of ultracold atoms}
\author{S. Mart\'\i nez-Garaot}
\affiliation{Departamento de Qu\'{\i}mica F\'{\i}sica, UPV/EHU, Apdo.
644, 48080 Bilbao, Spain}
\author{E. Torrontegui}
\affiliation{Departamento de Qu\'{\i}mica F\'{\i}sica, UPV/EHU, Apdo.
644, 48080 Bilbao, Spain}
\author{Xi Chen}
\affiliation{Department of Physics, Shanghai University, 200444
Shanghai, People's Republic of China}
\author{M. Modugno}
\affiliation{{Dpto.~de F\'isica Te\'orica e Hist.~de la Ciencia, Universidad del Pa\'is Vasco UPV/EHU, 48080 Bilbao, Spain}}
\affiliation{IKERBASQUE, Basque Foundation for Science, Alameda Urquijo 36, 48011 Bilbao, Spain}%
\author{D. Gu\'ery-Odelin}
\affiliation{Laboratoire de Collisions Agr\'egats R\'eactivit\'e,
CNRS UMR 5589, IRSAMC, Universit\'e de Toulouse (UPS), 118 Route de
Narbonne, 31062 Toulouse CEDEX 4, France}
\author{Shuo-Yen Tseng}
\affiliation{Department of Photonics, National Cheng Kung University, Tainan 701, Taiwan}
%
\author{J. G. Muga}
\affiliation{Departamento de Qu\'{\i}mica F\'{\i}sica, UPV/EHU, Apdo.
644, 48080 Bilbao, Spain}
\affiliation{Department of Physics, Shanghai University, 200444
Shanghai, People's Republic of China}
%
%
%
\begin{abstract}
Sending multiple messages on qubits encoded in different 
vibrational modes of cold atoms or ions along a transmission waveguide
requires to  merge first and then separate the modes at input and output ends.  
Similarly, different qubits can be stored in the modes of a trap and be separated later.     
We design the fast splitting of a harmonic trap into an asymmetric  double well so that  the initial ground vibrational state
becomes the ground state of one of two final wells, and the initial first excited state 
becomes the ground state of the other well.   
This might be done adiabatically by slowly  deforming the trap. We speed up the process by  
inverse engineering a  double-function trap using  dynamical invariants.   
The separation (demultiplexing) followed by an inversion of the asymmetric bias 
and then by the reverse process (multiplexing)  provides a population inversion protocol 
based solely on trap reshaping.    
\end{abstract}
%
%
\pacs{32.80.Qk, 37.10.Gh, 37.10.Vz, 03.75.Be}
%
\maketitle
%
%
%
%
%
%
%
%
%
{\it{Introduction.}--} One of the main goals of atomic physics is to achieve an exhaustive 
control of atomic states and dynamics \cite{advances}. The ultra-cold domain is particularly 
suitable for this aim as it provides a rich scenario of quantum states and phenomena. 
Atom optics and atomtronics \cite{atom} intend to manipulate cold atoms in circuits and devices 
for applications  in metrology, quantum information, or fundamental science. These devices are frequently
inspired by electronics (e.g. the atom diode \cite{ad1,at2}, the transistor \cite{atom}, atom chips \cite{chips}), 
or optics
(e.g. beam splitters \cite{bs}, or multiplexing \cite{optics1,optics2}).   

\begin{figure}[b]
\begin{center}
\hspace*{-.8cm}\includegraphics[height=1.3cm,angle=0]{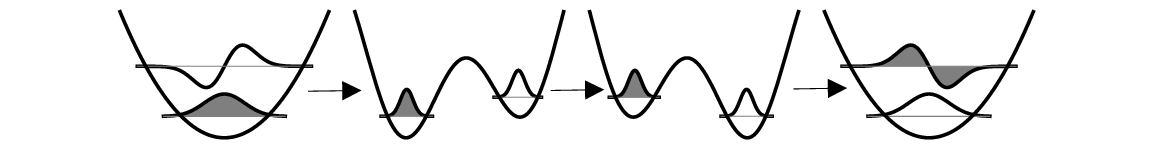}
\end{center}
\caption{\label{adiabatic1}
Population inversion using trap deformations in three steps:  
demultiplexing, bias inversion, and multiplexing.}
\end{figure}
%
In this paper we shall focus on a cold-atom realization of multiplexing, a basic process in modern telecommunications.    
Multiplexing is the transmission of different messages 
via a single physical medium. 
A multiplexer combines signals from several emitters into a single medium whereas a demultiplexer
performs the reverse operation. 
The concept of multiplexing is relevant for quantum information processing (for its use in quantum repeaters
see \cite{multip1,multip2}, or for trapped ions \cite{multip3}).  
We envision here optical or magnetic waveguides for atoms holding several transverse orthogonal modes \cite{ion,wg1,wg2,wg3}. 
If the qubit is encoded in the internal state of the atom, several qubits may be carried out simultaneously by different modes.   
To develop such a quantum-information  architecture, fast  multiplexers/demultiplexers that could  join
the modes from different waveguides into one guide, or separate them, are needed.
We shall discuss trap  designs for demultiplexing since the multiplexer would simply operate in reverse. 
For a proof-of-principle we propose the  simplified setting of a single initial harmonic trap 
for non-interacting cold atoms whose first two 
eigenstates will be separated, as in the first step of Fig. \ref{adiabatic1}, into two different wells.   
In a complete demultiplexing process the final wells should be independent, with negligible tunneling.   
The challenge  is to design the splitting   (a) without final excitation of higher vibrational levels, (b) 
in a short time, and (c) with a 
realizable trap potential.     
Condition (a) may be achieved by an adiabatic asymmetric splitting \cite{Gea,split} in
which, for  moderate bias compared to the vibrational quanta, the initial ground state becomes the ground state of the well with the 
lowest energy, and the excited state becomes the ground state of the other well. 
This adiabatic approach generally fails to satisfy the condition (b) which we shall implement 
applying  ``shortcuts to adiabaticity'' \cite{Chen,ChenPRA,review,Bowler}. As for (c), we shall make use of a simple two-level model for the shortcut design,
and then map it to a realistic potential
recently 
implemented to  realize an atomic Josephson junction 
\cite{Oberthaler}.
Finally, several applications, such as separation of multiple modes, population inversion, or controlled excitation,  will be discussed.

{\it{Slow adiabatic and fast adiabatic processes}.--}
Suppose that a harmonic potential evolves adiabatically into two well-separated and asymmetric wells as in 
the first step  of Fig. \ref{adiabatic1}. 
To accelerate the dynamics we shall use a  moving two-level approximation based on a 
(yet-unspecified) process where  
a symmetrical potential evolves from an initial harmonic trap  to a final double well.  
Then, we construct a time-dependent
orthogonal bare basis 
$|L(t)\rangle = \left(\scriptsize{\begin{array} {rccl} 0\\ 1 \end{array}} \right)$, $|R(t)\rangle = \left(\scriptsize{\begin{array} {rccl} 1\\ 0 \end{array}} \right)$ of left and right states, obtained by a
linear combination of the instantaneous ground and first excited
states. 
An approximate two-mode Hamiltonian model for a generally  {\it{asymmetrical}} process is written in this basis as 
\beq
\label{H_tm}
H_{2\times2}(t)=\frac{\hbar}{2} \left ( \begin{array}{cc}
\lambda(t)
& -\delta(t)\\
-\delta(t)& -\lambda(t)
\end{array} \right),
\eeq
where, for the double well  configuration, $\delta(t)$ is  the tunneling rate, and $\hbar\lambda(t)$ the relative gap, or bias, between the two wells. 
For the initial harmonic potential at $t=0$, $\lambda(0)=0$ and $\delta(0)=\omega_0$.  
The instantaneous eigenvalues are
%
$E^{\pm}_\lambda(t)=
\pm \frac{\hbar}{2} \sqrt{\lambda^2(t)+\delta^2(t)}$,
%
and the normalized eigenstates 
\beqa
\label{eigenstates_tls}
|\psi^+_\lambda(t)\rangle &=& \sin{ \left ( \frac{\alpha}{2} \right ) } |L(t)\rangle-\cos {\left ( \frac{\alpha}{2} \right ) }|R(t)\rangle, 
\\
|\psi^-_\lambda(t)\rangle &=& \cos{ \left ( \frac{\alpha}{2} \right )}|L(t)\rangle+\sin{\left ( \frac{\alpha}{2} \right )}|R(t)\rangle,
\nonumber
\eeqa
where the mixing angle $\alpha=\alpha(t)$ is given by $\tan \alpha = \delta (t)/\lambda(t)$. 
The boundary conditions on $\lambda(t)$ and $\delta(t)$ are
\beq
\label{bc_d}
\delta(0)=\omega_0, \, \,   \lambda(0)=0,\,\,
\delta(t_f)=0, \, \,   \lambda(t_f)=\lambda_f, 
\eeq
which correspond, at time $t=0$, to a harmonic well, and at time $t_f$ to two independent wells
with asymmetry bias $\hbar\lambda_f$.   

To design a fast, but still adiabatic process, we shall first assume the simplifying conditions: $\lambda(t)=\lambda$ constant and
$\lambda/\delta(0)\ll 1$. Thus $\alpha(0)\approx\pi/2$ and the initial eigenstates essentially coincide with the ground and 
first excited states of the harmonic oscillator.
For a constant $\lambda$ the adiabaticity condition  reads \cite{split}
$\left |\frac{\lambda \dot \delta(t)}{2(\lambda^2+\delta(t)^2)^{3/2}} \right |\ll1$. 
Imposing a constant value $c$ for the adiabaticity parameter 
and using the boundary conditions 
for $\delta$ in 
Eq. (\ref{bc_d}), we  fix the integration constant 
and the value of $c$,
%
$c=\frac{\omega_{0}}{2 \lambda \sqrt{\omega_{0}^2+\lambda^2}\, t_f}$. 
%
The ``fast adiabatic'' solution of the differential equation for $\delta(t)$ takes finally the form 
%
$\delta_{fa}(t)=\frac{\omega_0\lambda(t_f-t)}{\sqrt{{\lambda^2t_f^2+\omega_0^2t(2t_f-t) }}}$.
%
However, to keep adiabaticity,  $c\ll 1$ should hold, so this protocol is limited 
by  
$t_f\gg \frac{1}{2\lambda}\gg\frac{1}{\omega_0}$.
%
We shall now work out an alternative, faster protocol
based on invariants in which the boundary conditions
on $\lambda(t)$ and $\delta(t)$  
will be satisfied exactly.     

{\it{Invariant-based inverse engineering}.--}
For the Hamiltonian in Eq. (\ref{H_tm})
there is a dynamical invariant $I(t)$  of the form \cite{ChenPRA}
\beq
\label{inv}
I(t)=\frac{\hbar}{2}\Omega_0 \left ( \begin{array}{cc}
\cos \theta(t)
& \sin \theta(t)e^{i\varphi(t)}\\
\sin \theta(t)e^{-i\varphi(t)}& -\cos \theta(t)
\end{array} \right),
\eeq
where $\varphi(t)$ and $\theta(t)$ are auxiliary (azymuthal and polar) angles, and
$\Omega_0$ is an arbitrary constant with units of frequency. 
The eigenvectors of $I(t)$
multiplied by Lewis-Riesenfeld phase factors
provide two orthogonal solutions of the time-dependent Schr\"odinger equation \cite{ChenPRA}.  
To inverse engineer the Hamiltonian we design the invariant 
first, and then deduce the Hamiltonian from it. 
The boundary conditions $[H_{2\times2}(t),I(t)]=0$ will be applied at the interval ends $t_b=0,\,t_f$, so that the eigenvectors
of $I(t_b)$ and $H_{2\times2}(t_b)$ coincide.  The role of the invariant is therefore to drive the initial eigenstates 
of $H_{2\times2}(0)$  to the eigenstates of $H_{2\times2}(t_f)$. In  our application this implies a unitary mapping
from the first two eigenstates of the harmonic oscillator to the ground states of the left and right final wells.      

From the invariance property 
$i\hbar\frac{\partial I(t)}{\partial t}- [H_{2\times2}(t),I(t)]=0$,
%
it follows that 
\beqa
\label{inv_de}
\delta(t)&=&-\dot\theta(t)/\sin{\varphi(t)},
\\ \nonumber
\lambda(t)&=&-\delta(t)\cot{\theta(t)}\cos{\varphi(t)}-\dot \varphi(t).
\eeqa
The commutativity of $I(t)$ and $H_{2\times2}(t)$ at boundary times $t_b=0,t_f$ imposes  the conditions
\beq
\label{con_0}
\begin{array}{ll}
\lambda(t_b) \sin[\theta(t_b)]e^{i\varphi(t_b)}+\delta(t_b)\cos[\theta(t_b)] = 0,
\\
\lambda(t_b) \sin[\theta(t_b)]e^{-i\varphi(t_b)}+\delta(t_b)\cos[\theta(t_b)] = 0,
\\
\delta(t_b)\sin[\theta(t_b)]\sin[\varphi(t_b)]=0,
\end{array}
\eeq
Taking into account Eqs. (\ref{bc_d})
we get from Eq. (\ref{con_0})
\beq
\label{b_c}
\theta(0)=\pi/2,\;\;
\varphi(0)=\pi,\;\;
\theta(t_f)=\pi,\;\;
\dot\theta(t_f)=0.
\eeq
%
%
%
\begin{figure}[b]
\begin{center}
\includegraphics[height=2.6cm,angle=0]{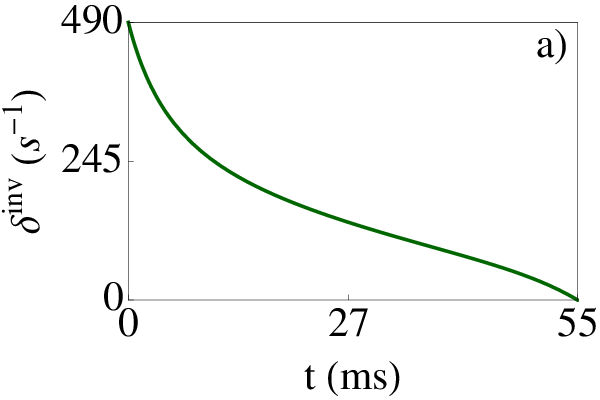}
\includegraphics[height=2.6cm,angle=0]{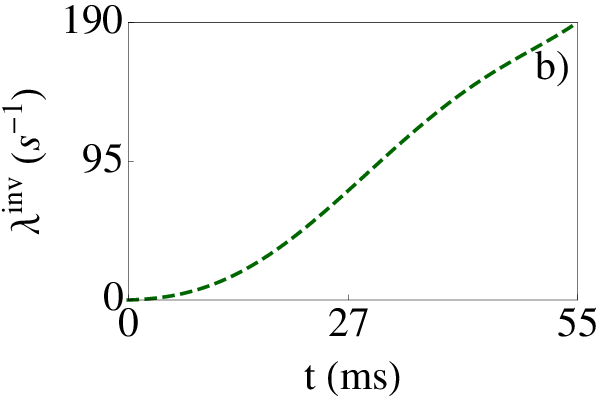}
\end{center}
\caption{\label{deltainv_10}
(Color online)
(a) $\delta^{\rm{inv}}(t)$ and (b) $\lambda^{\rm{inv}}(t)$.
$\delta(0)=2\pi\times 78$ Hz, 
$\lambda_f=190$ s$^{-1}$, $\dot\lambda(0)=190$ s$^{-2}$, and $t_f=55$ ms.}
\end{figure}
%
These conditions lead to indeterminacies in Eq. (\ref{inv_de}). To resolve them we apply L'H\^ opital's rule repeatedly and find  
additional boundary conditions,
\beqa
\label{b_c_2}
\dot\theta(0)\!&\!=\!&\ddot\theta(0)=\dot{\varphi}(0)=0,\, \, \dddot\theta(0)=-\omega_0\dot\lambda(0),
\\\nonumber
\ddot\varphi(0)\!&\!=\!&-\dot\lambda(0),\,\, 
\varphi(t_f)=\pi/2,\;\;
\dot\varphi(t_f)=-\frac{\lambda_f}{3}.
\eeqa
with $\dot\lambda(0)\neq 0$.
At intermediate times, we interpolate the angles assuming a polynomial ansatz, 
$\theta(t)=\sum_{j=0}^5 a_j t^j$ and 
$\varphi(t)=\sum_{j=0}^4 b_j t^j$, where the
coefficients are found by solving the equations for the boundary conditions. 
Thus we obtain  the Hamiltonian functions $\delta^{\rm{inv}}(t)$ 
and $\lambda^{\rm{inv}}(t)$ from Eq. (\ref{inv_de}).  
Figure \ref{deltainv_10} 
provides an example of parameter trajectories.

\begin{figure}[b]
\begin{center}
\includegraphics[height=2.6cm,angle=0]{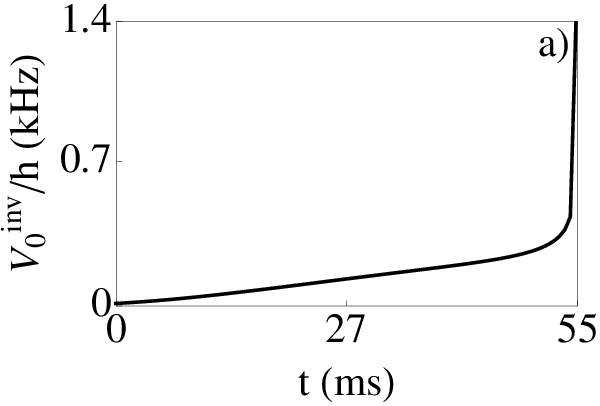}
\includegraphics[height=2.6cm,angle=0]{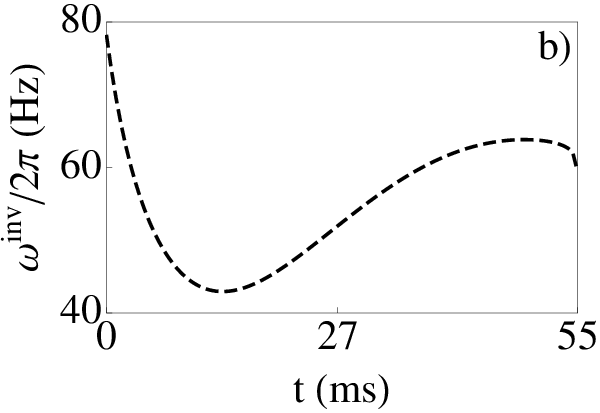}
\end{center}
\caption{\label{v0inv_10}
Lattice height $V_0$, and trap frequency $\omega/(2\pi)$ using invariant-based engineering and mapping.    
$\Delta x=200$ nm.  
}
\end{figure}

{\it{Mapping to coordinate space\label{mapp}}.--}
Our purpose now is to map the $2\times2$ Hamiltonian into a
realizable potential 
\beq
\label{V_Oberthaler}
V(x,t)=\frac{1}{2}m\omega^2x^2+V_0\cos^2 \left [ \frac{\pi(x-\Delta x)}{d_l} \right ]. 
\eeq
This form has been already implemented \cite{Oberthaler} with optical dipole potentials, combining a harmonic confinement 
due to a crossed beam dipole trap with a periodic light shift potential provided by the interference pattern of two mutually coherent laser beams.  
The control parameters are in principle the frequency $\omega$, the displacement $\Delta x$ of the optical lattice relative 
to the center of the harmonic well, the amplitude $V_0$, and the lattice constant $d_l$,  
but in the following examples we fix $d_l$ and $\Delta x$;  the other two parameters offer enough flexibility and 
are easier to control as time-dependent functions.     
To perform the mapping, we minimize numerically 
%
$F[V_0(t),\omega(t)]=[\delta^{\rm{id}}(t)-\delta(t)]^2+[\lambda^{\rm{id}}(t)-\lambda(t)]^2$,
%
using the simplex method.  
The functions $\delta^{\rm{id}}(t)$ and $\lambda^{\rm{id}}(t)$ are designed according to the shortcuts 
discussed before, 
and 
$\delta(t)$ and $\lambda(t)$ are computed  
as 
%
$\delta(t)=-\frac{2}{\hbar}\langle L(t)|H|R(t) \rangle =-\frac{2}{\hbar}\langle R(t)|H|L(t) \rangle$,
$\lambda(t)=\frac{2}{\hbar}\langle R(t)|H-\Lambda|R(t) \rangle =-\frac{2}{\hbar}\langle L(t)|H-\Lambda|L(t) \rangle$,
%
where $H=H(V_0(t),\omega(t); \Delta x, d_l)=-\frac{\hbar^2}{2m}\frac{\partial^2}{\partial x^2}+V$ is the full Hamiltonian in coordinate space 
with a kinetic energy term and the potential (\ref{V_Oberthaler});  $\Lambda(t)=[E_{\lambda}^-(t)+E_{\lambda}^+(t)]/2$ is a shift 
defined from the first two levels $E_\mp$ of $H$ to match the zero-energy point between the coordinate and the two-level system;  
finally, $|R(t)\ra=(|g(t)\ra+|e(t)\ra)/2^{1/2}$ and $|L(t)\ra=(|g(t)\ra-|e(t)\ra)/2^{1/2}$ form the base, where  
$|g(t)\ra$ is the ground state and  $|e(t)\ra$ the first excited state  
of the symmetrical Hamiltonian   $H_0(V_0(t),\omega(t); \Delta x=0, d_l)$, defined as $H$ but with $\Delta x=0$, 
which we diagonalize numerically.       
In our calculations, $\delta(t)$ and $\lambda(t)$ become indistinguishable from their 
ideal counterparts.   Figure \ref{v0inv_10}  depicts   $V_0(t)$ and $\omega(t)$ for the  parameters   
of Fig. \ref{deltainv_10}. We use $^{87}$Rb atoms  and a lattice spacing $d_l=5.18$ $\mu$m. 
The sharp final increase of $V_0(t)$  makes the two wells totally independent but,    
for most applications this strict condition may be relaxed to avoid  intra-well excitations. 

\begin{figure}[h]
\begin{center}
\includegraphics[height=2.6cm,angle=0]{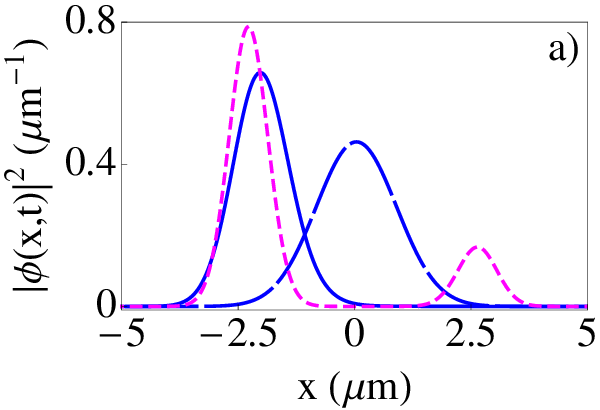}
\includegraphics[height=2.6cm,angle=0]{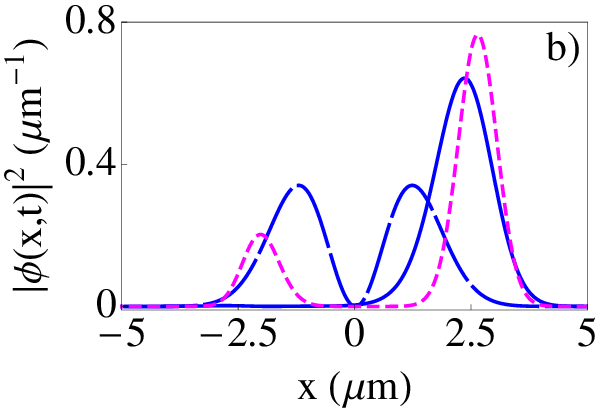}
\end{center}
\caption{\label{psiinv_10}
(Color online)
(a): Ground state at $t=0$ (long-dashed,  blue line); final state with the shortcut  (solid, blue line, indistinguishable 
from the ground state of the final trap); final state with linear ramp for $V_0(t)$
and $\omega=2\pi\times78$ Hz (short-dashed,  magenta line).  
(b): Same as (a) for the first excited state. 
Parameters like in Fig. \ref{v0inv_10} at $t=53$ ms. 
The linear ramp for $V_0(t)$ ends in the same value 
used for the shortcut.} 
\end{figure}

\begin{figure}[h]
\begin{center}
\includegraphics[height=2.6cm,angle=0]{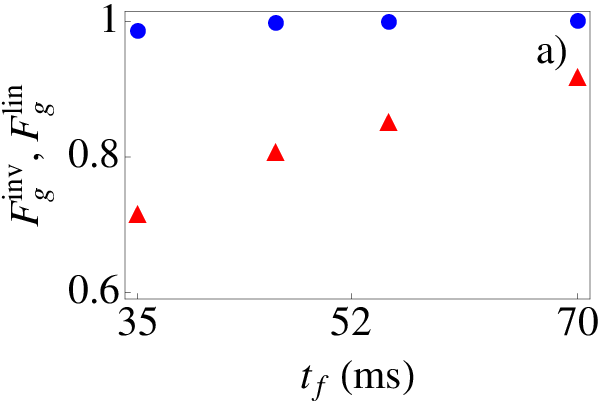}
\includegraphics[height=2.6cm,angle=0]{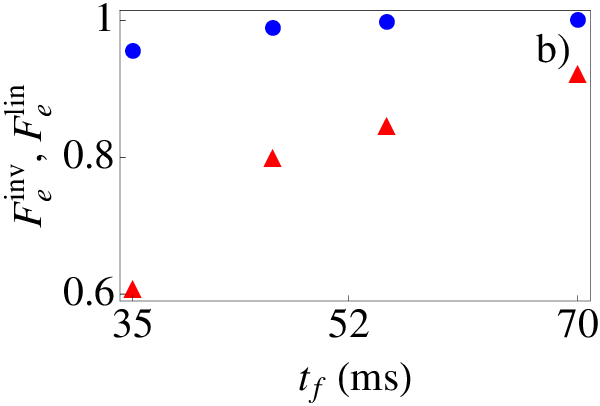}
\end{center}
\caption{\label{Fidelity}
(Color online)
Fidelities with respect to the final ground state starting at the ground state (a) and with respect to the final first excited state starting at the excited state (b) versus final time $t_f$, 
via shortcuts  ($F_{g}^{inv}$ and $F_{e}^{inv}$, blue circles), or  linear ramping  of $V_0(t)$ ($F_{g}^{lin}$ and $F_{e}^{lin}$, red triangles). 
The fidelity is computed at 2 ms less than the nominal $t_f$. Other parameters as in Figs. \ref{deltainv_10}, \ref{v0inv_10}, and \ref{psiinv_10}.
}
\end{figure}

\begin{figure}[h]
\begin{center}
\includegraphics[height=2.6cm,angle=0]{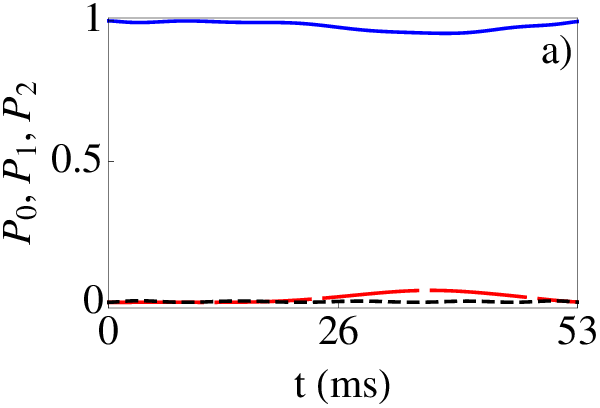}
\includegraphics[height=2.6cm,angle=0]{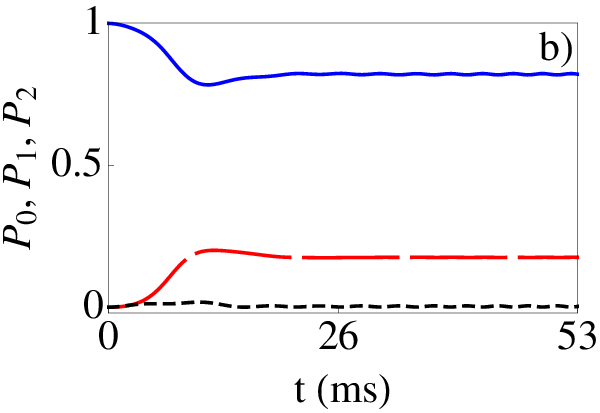}
\end{center}
\caption{\label{pinv_10}
(Color online)
Populations of the states for the shortcuts (a) and the linear ramp for $V_0(t)$ (b). 
Ground state ($P_0$, solid blue line); First excited state ($P_1$, long-dashed red line); Second excited state ($P_2$, short-dashed black line). 
Parameters as in Fig. \ref{psiinv_10} (a).
}
\end{figure}

%
Figure \ref{psiinv_10} demonstrates perfect transfer for the ground (a) and the excited state (b)
using the very same protocol in both cases, the one depicted in Figs. \ref{deltainv_10} and Ê\ref{v0inv_10}.  
(Thanks to the superposition principle, the
same protocol would produce a perfect demultiplexing for any linear
combination of the ground and excited states.) Initial and final states are represented, solving the Schr\"odinger equation with the potential (\ref{V_Oberthaler}).
We stop the process $2$ ms before the nominal time $t_f$ as the fidelity
reaches a stable maximum there and a further increase of $V_0$ is not required. 
We also include the  results for the protocol in which $\omega$
is kept constant and $V_0(t)$ is a linear ramp (with the same durations as the  shortcut protocols). For    
this linear protocol the final state includes a significant  density in the ``wrong'' well.   
This simple linear-$V_0$  approach needs $t_f\gtrsim0.7$ s  to become adiabatic and produce  the same fidelity, 
0.9997,  found for a shortcut protocol ten times faster, $t_f=0.07$ s, the rightmost point in Fig. \ref{Fidelity} (a).    
Fig. \ref{pinv_10} compares  the populations   
in the instantaneous basis of the (full, coordinate-space) Hamiltonian for the shortcut and the linear protocols
when the system starts in the ground state, corresponding to  Fig. \ref{psiinv_10} (a). 
The shortcut protocol implies a transient exchange between ground and (first) excited levels but finally 
takes the system to the desired ground state. In contrast for the linear protocol the excitation is permanent
leading to a poor final fidelity. 

In the two-level model 
$t_f$ may be reduced arbitrarily,  but in the coordinate space Hamiltonian levels 0-1 will only be ``independent'' as long as 
higher levels are not excited.  These excitations are the limiting factor to shorten the times further with the 
current mapping scheme. Some guidance is provided 
by the Anandan-Aharonov relation  $t_f>h/(4 \overline{\Delta E})$, where $\overline{\Delta E}$ is the time average of the
standard deviation \cite{AA}.   
%

%
%

%
%
%
%
%
%
%

{\it{Discussion.}--}
Vibrational multiplexing may be combined with internal-state multiplexing \cite{multip3} to provide 
a plethora  of possible operations. Motivated by the prospective use of multiplexing/demultiplexing  for quantum information 
processing,  we have applied shortcuts-to-adiabaticity techniques to speed up the spatial separation of vibrational modes 
of a harmonic trap.  
A  similar approach 
would separate $n$ modes  into $n$ wells so as to deliver more information into different processing sites. 
The number of modes that could be separated 
will depend on the asymmetric bias in relation to other potential parameters:  
the bias among the extreme wells should not exceed the
vibrational quanta in the final wells. 
The bias determines possible speeds too, as smaller biases generally imply longer times.  
           
A previous work \cite{split} dealt also with splitting operations and shortcuts to adiabaticity, 
but the objective was the opposite to our aim here. 
Since adiabatic following from a harmonic trap to an asymmetric  double well collapses the ground state wave to one of the two wells, 
a  ``fast-forward''  (FF) technique  \cite{Masuda,ourff}  was applied to {\em avoid} the collapse and achieve perfect, 
balanced  splitting, as required, e.g., for matter-wave interferometry.  
The idea was that for a fast non-adiabatic shortcut the perturbative effect of the asymmetry becomes negligible. 
The stabilizing effect of interactions was also
characterized within a mean-field treatment.  
In the present paper the objective is to send each mode of the initial harmonic trap as fast as possible 
to a different final well, so we needed  a different methodology.  
Instead of FF, which demands an arbitrary control of the potential function 
in position and time, we have restricted the potential to a form with a few controllable
parameters (in practice we have let only two of them evolve in time). Inverse engineering of the Hamiltonian is carried
out for a two-level model using invariants of motion, and the resulting (analytical) Hamiltonian is then mapped 
to real space.  The
discrete Hamiltonian
is useful  as it provides a simple picture to understand and design the dynamics at will. In future work we shall increase the number of  
levels in the discrete model and test alternative potential functions.  
The method provides also a good basis to apply optimal control theory (OCT), which complements 
invariant-based engineering, see. e.g. \cite{Li},  
by selecting among the 
fidelity-one protocols
according to other physical requisites.  
As for interactions and nonlinearities,  they will generally spoil a clean multiplexing/demultiplexing
processes, so we have only examined linear dynamics here.    
     
An application of the demultiplexing schemes discussed in this work is the population inversion of the first two levels of the harmonic trap 
without making use of internal state excitations \cite{Salomon}.  This is useful to avoid decoherence effects induced by decay, 
or for species without an appropriate (isolated two-level)  structure. 
The scheme is based on the three steps shown in Fig. \ref{adiabatic1}. 
A mechanical excitation of the ground state level into the first excited state of a fixed  anharmonic potential 
was implemented by shaking the trap along a trajectory calculated with an OCT algorithm \cite{ms_schmiedmayer}.  
Our proposed approach  relies instead on a smooth potential deformation. This type of inversion could be applied to interacting
Bose-Einstein condensates
as long as the initial states are pure ground or excited levels. 
The production of twin-atom beams from the excited state is an outstanding application \cite{twin}. 

Asymmetric double wells may also be used for other state-control operations such as preparing non-equilibrium Fock states 
through a ladder excitation process. The vibrational number may be increased by one at every step. 
Each excitation would start and finish with  demultiplexing and multiplexing 
operations from the harmonic oscillator to the double well and viceversa,  as described in the main text. 
Between them  the two wells are independent and 
their height or width can be adjusted to produce the desired level ordering. 
For an even-to-odd vibrational number transition 
this requires an inversion of the bias, as in Fig. \ref{adiabatic1};
transitions from odd to even levels  are performed by deepening the left well until the initially occupied level on the right well  
surpasses one of the levels in the left  well. The steps may be repeated until a given Fock state is reached.  
Operating in reverse mode, a given excited state could be taken down to the ground state, as in sideband cooling, 
just with trap deformations.

Open questions left for future work include optimizing 
the robustness of parameter trajectories versus noise and perturbations \cite{Andreas}, 
or finding time bounds in terms of average energies, similar to the ones 
for harmonic trap expansions \cite{energy} or transport
\cite{transport}. 
The present results may also be applied 
for  optical waveguide design \cite{SCh}, or to 2D systems as a way to generate vortices.       

{\it{Acknowledgment.}--}
This work was supported by the National Natural Science Foundation
of China (Grant No. 61176118), 
the Grants
No. 12QH1400800
IT472-10, BFI-2010-255, 13PJ1403000,
FIS2012-36673-C03-01, and the program UFI 11/55.
S. M.-G. acknowledges a fellowship by UPV/EHU.  
%
%
%

\end{document}